\documentclass{PoS}

\title{Testing the sterile neutrino dark matter paradigm with astrophysical observations}

\ShortTitle{Structure formation and sterile neutrino dark matter}

\author{\speaker{Aurel Schneider}\\
        Institute for Astronomy, Department of Physics, ETH Zurich, Wolfgang-Pauli-Strasse 27, 8093, Zurich, Switzerland\\
        E-mail: \email{aurel.schneider@phys.ethz.ch}}

\abstract{Sterile neutrino dark matter is expected to suppress structure formation at small astrophysical scales. The details of the suppression depend on the sterile neutrino production mechanism in the early universe. In this proceeding, we focus on the most popular cases of resonant production (via the mixing between active and sterile neutrinos) and scalar decay production (via the decay of a hypothetical scalar singlet). We first review current constraints from structure formation before discussing how the sterile neutrino dark matter hypothesis can alleviate the overabundance problem of dwarf galaxies in the local universe.}

\FullConference{Neutrino Oscillation Workshop\\
		4 - 11 September, 2016\\
		Otranto (Lecce, Italy)}

\begin{document}
\section{Introduction}
Sterile neutrinos are considered an attractive dark matter (DM) candidate because they only require a minimal, well motivated extension to the standard model of particle physics, and they lead to suppressed clustering at the scales of dwarf galaxies. The latter potentially solves long-standing puzzles of small-scale structure formation in astrophysics, such as the \emph{over-abundance} \cite{Klypin1999,Moore1999} or the \emph{cusp-core}  \cite{deBlok2010} problems.

The sterile neutrino production mechanism in the early universe has a direct effect on the growth of matter perturbations which is governed by both the DM particle mass and the momentum distribution. In this proceeding, we focus on two of the most popular mechanisms: sterile neutrino production via resonantly enhanced mixing with active neutrinos (i.e. resonant production \cite{Shi1999}) and production via the decay of a hypothetical Higgs-like scalar singlet (i.e. scalar decay production \cite{Kusenko2006,Shaposhnikov2006}). The first mechanism relies on the mixing angles between active and sterile species and requires a strong degree of lepton asymmetry, the second mechanism depends on the couplings of the scalar singlet with the sterile neutrino and the standard model Higgs particle \cite{Koenig2016}.

In section~\ref{sec:constraints} we review current constraints from structure formation on the parameter space of the two production mechanisms and we discuss the viability of these constraints. In section~\ref{sec:VF} we show that sterile neutrino DM models which pass these constraints are still able to solve inconsistencies related to small-scale structure formation.

\section{Constraints from structure formation}\label{sec:constraints}
The particle mass of sterile neutrino DM is restricted to the \emph{keV} range, since sterile neutrinos would otherwise either decay too quickly or not cluster enough. The decay rate increases with particle mass, yielding an excess of X-ray photons that are potentially observable. The suppression of matter perturbations, on the other hand, becomes stronger with decreasing particle mass, resulting in fewer small galaxies. Both effects provide independent limits which can be used to efficiently constrain sterile neutrino DM. 

\begin{figure*}
\center{
\includegraphics[width=1.0\textwidth,trim={0.4cm 2.5cm 4.5cm 0.0cm}]{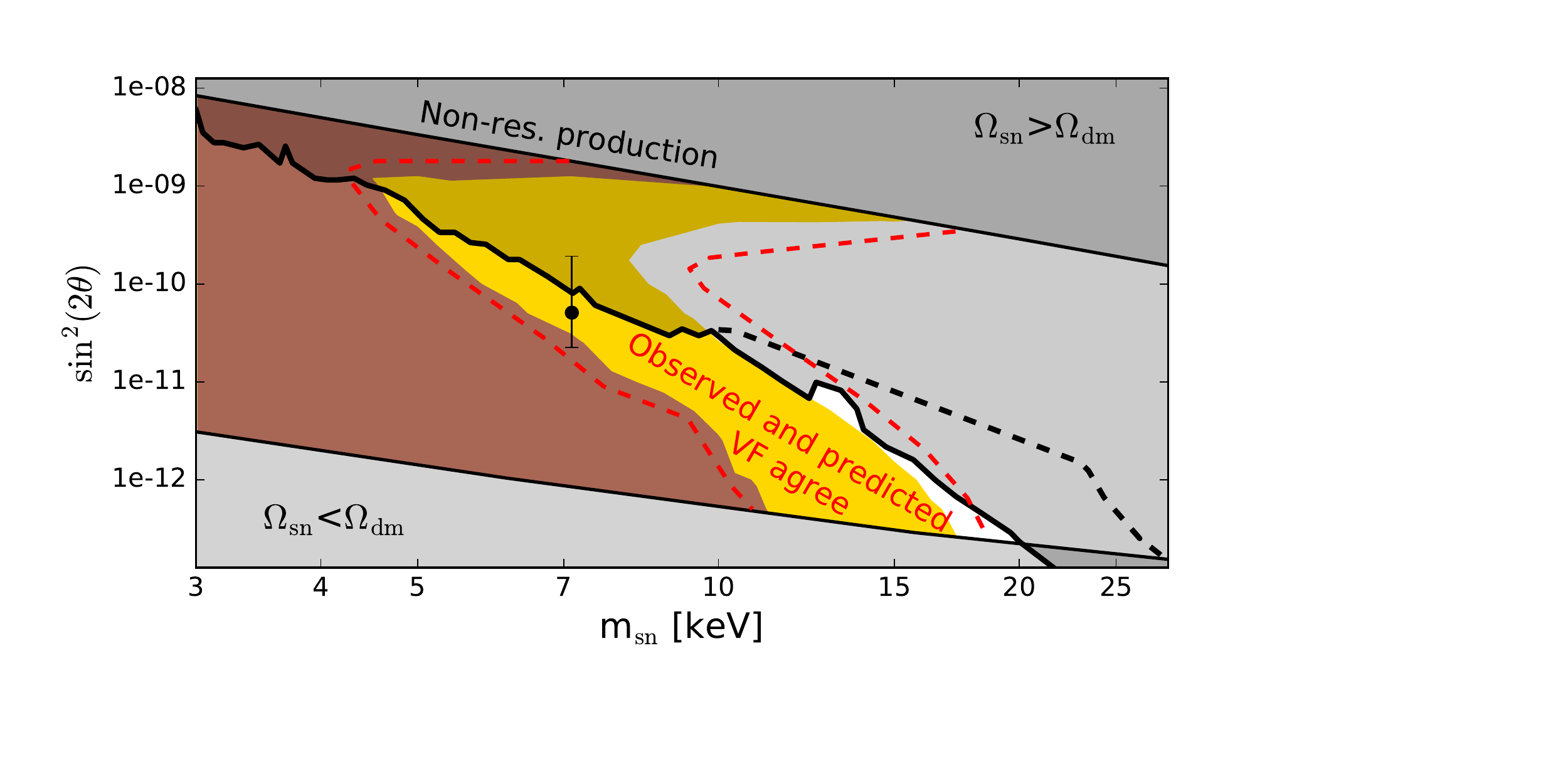}
\caption{\label{fig:resprod}Parameter space of resonantly produced sterile neutrino dark matter. Upper and lower grey areas indicate regions of DM over- and under-production. The X-ray limit from the non-observation of sterile neutrino decay is shown as thick black line (a more conservative limit is given by the dashed black line). Exclusion areas from structure formation based on conservative and progressive constraints are shown in brown and yellow, respectively. Models with parameters between the magenta dashed lines provide a solution to the dearth of dwarf galaxies in the nearby universe.}}
\end{figure*}

We first focus on the resonant production scenario and discuss constraints from structure formation originally presented in Ref.~\cite{Schneider2016a}. Fig.~\ref{fig:resprod} illustrates the parameter space of particle mass versus mixing angle for resonantly produced sterile neutrinos. The broad band between the upper and lower dark-grey areas (delimited by thin black lines) corresponds to the parameter space where the right DM abundance ($\Omega_{\rm dm}$) can be obtained by adjusting the lepton asymmetry accordingly. The upper thin-line represents the special case of non-resonant production (with zero lepton asymmetry). The lower thin line corresponds to the limit where the lepton asymmetry is too high to be in agreement with primordial nucleosynthesis. 

A compilation of X-ray limits (from the non-observation of photons due to sterile neutrino decay) is shown as thick solid line in Fig.~\ref{fig:resprod} constraining the parameter-space from the top-right (the dashed black line gives an alternative limit based on different observations and more conservative assumptions). For more information, see Ref.~\cite{Schneider2016a}.

Astrophysical observations, while being very sensitive to sterile neutrino production, might suffer from astrophysical uncertainties such as unknown baryon effects or observational systematics. We therefore give both a conservative and a more progressive limit. The conservative limit is shown as brown exclusion area in Fig.~\ref{fig:resprod}. It is in agreement with various constraints from Milky-Way satellite counts, high-redshift galaxies, or the Lyman-$\alpha$ forest using the most conservative assumptions. The progressive limit is shown as a yellow exclusion area. It is based on the Lyman-$\alpha$ constraints from Ref.~\cite{Viel2013} and adopted to resonant sterile neutrino DM by Ref.~\cite{Schneider2016a}. The limits from structure formation combined with the X-ray observations strongly reduce the allowed parameter space, leaving little room for the resonant production scenario.

\begin{figure*}
\center{
\includegraphics[width=1.0\textwidth,trim={0.4cm 2.5cm 4.5cm 0.0cm}]{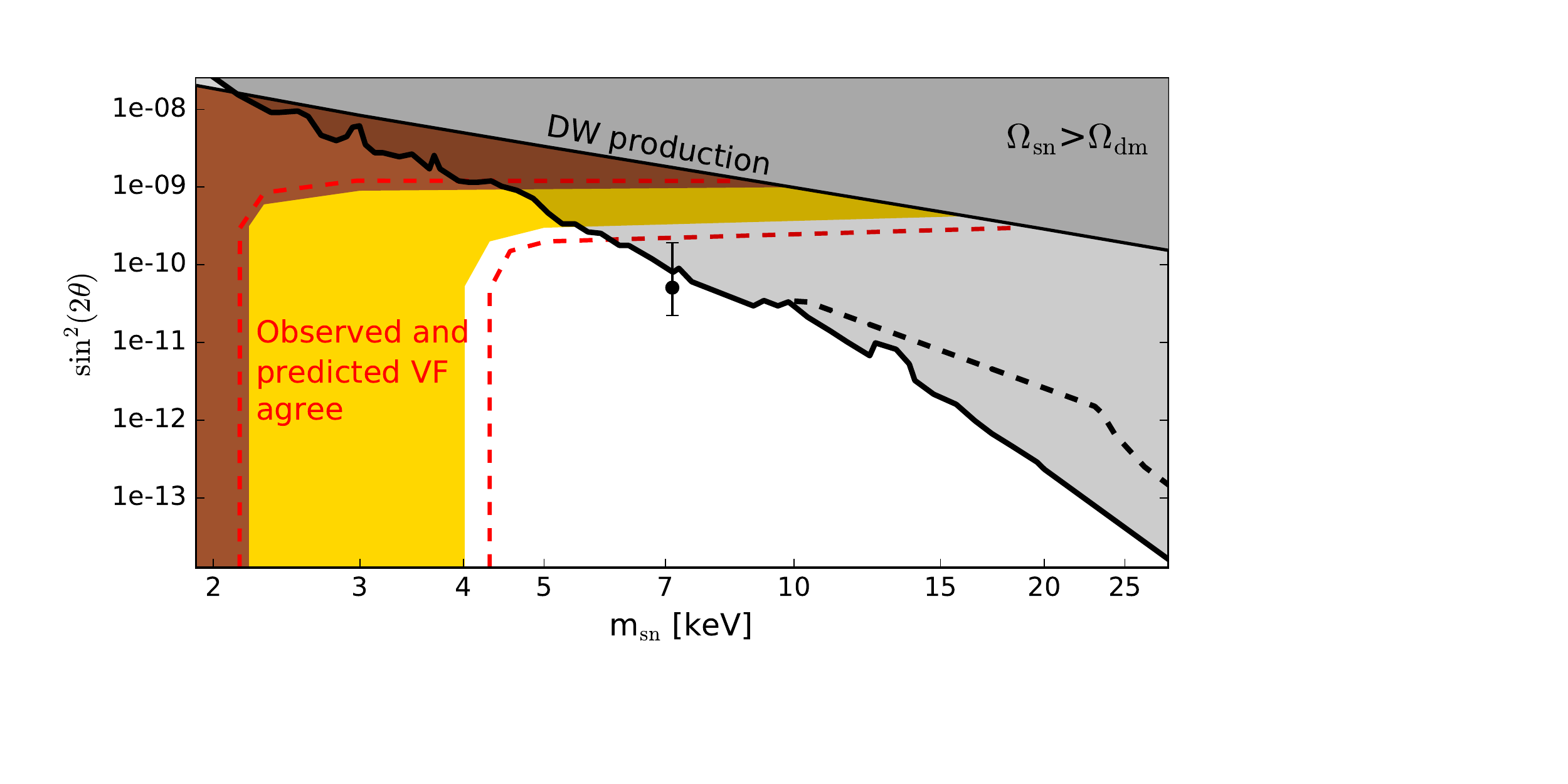}
\caption{\label{fig:SDprod}Same than Fig.~1 but for scalar decay production.}}
\end{figure*}

In a next step we focus on sterile neutrino dark matter produced by the decay of a hypothetical scalar singlet. The available parameter space is illustrated in Fig.~\ref{fig:SDprod}. In the most optimal scenario of a very weak coupling between the scalar and the standard model, the resulting sterile neutrino momenta are very small \cite{Koenig2016}. This considerably weakens the limits from structure formation, as shown by the conservative and progressive exclusion areas in Fig.~\ref{fig:SDprod} (brown and yellow). Since the scalar decay production mechanisms is not tied to the mixing angle, the available parameter space extends to arbitrarily small mixing angles, resulting in the absence of a lower limit in Fig.~\ref{fig:SDprod}.

\section{The sterile neutrino solution to the galactic velocity function}\label{sec:VF}
The number density of galaxies as a function of rotational velocity -- the galactic velocity function (VF) -- combines the abundance of dwarf galaxies with their inner structure, making it a particularly interesting statistic of small-scale structure formation. It has been shown recently that the observed VF is in disagreement with expectations from the $\Lambda$CDM model \cite{Klypin2015} and that this disagreement is unlikely solvable with strong baryonic effects \cite{Trujillo-Gomez2016}\footnote{Several studies based on simulations have shown to be able to solve the galactic VF with strong feedback mechanisms that make small galaxies reside in large haloes \cite{Maccio2016,Brooks2017}. However, this comes at the prize of not reproducing the galaxy-halo relation inferred from direct mass estimates based on rotation curves of field dwarfs (see \cite{Trujillo-Gomez2016,Schneider2016b} for a detailed discussion).}. On the other hand, alternative DM models  -- such as sterile neutrinos -- might provide natural solutions to the observed discrepancy \cite{Schneider2016b}.

We use the method developed in Ref.~\cite{Schneider2016b,Schneider2015} to determine the parts of the parameter space of sterile neutrino DM, where the discrepancy between the observed and the predicted VF disappears (i.e. where observations and the model predictions overlap). The allowed region is delimited by magenta dashed lines in Fig.~\ref{fig:resprod} and Fig.~\ref{fig:SDprod} for the case of resonant production and scalar decay production, respectively. It includes parts of the parameter space that is allowed by both the conservative and the progressive constraints from structure formation. This means that there are specific models of sterile neutrino DM which are able to solve small-scale problems of structure formation while being in agreement with all current constraints.

\end{document}